\documentclass[11pt]{amsart}
\usepackage{amsmath,amsfonts,eucal,amssymb}
\usepackage[retainorgcmds]{IEEEtrantools}	
\usepackage{verbatim}	
\usepackage[all]{xy}
\usepackage{tikz}

\title[Simulation Quantization]{Simulation Quantization:\\ An Application of Physical Versions of the Church-Turing Thesis}
\author{Abel Wolman}
\address{The Johns Hopkins University,
Baltimore, Maryland 21218}
\email{awolman2@jhu.edu}
\subjclass{}
\date{December 6, 2019}

\begin{document}

\allowdisplaybreaks

\newcommand{\cal}{\mathcal}
\newcommand{\cX}{{\mathcal X}}
\newcommand{\cZ}{{\mathcal Z}}
\newcommand{\Op}{{\rm op}}
\newcommand{\onto}{{\rm onto}}
\newcommand{\rv}[1]{\mathbf{#1}}
\newcommand{\bb}[1]{\mathbb{#1}}
\newcommand\noi{\noindent}

\newtheorem{thm}{Theorem}

\hyphenation{quasi-math-e-ma-ti-cal i-so-chro-nous cor-res-pon-dence-prin-ciple}

\begin{abstract}
\noindent
An integrable anharmonic oscillator is presumably simulable by a classical computer and therefore by a quantum computer. An integrable anharmonic oscillator whose Hamiltonian is of normal type and quartic in the canonical coordinates is not quantizable. If, as argued here, quantum simulation of a finitely realizable classical physical system entails quantization of that system, then either there exist nonsimulable, integrable anharmonic oscillators or there are no obstructions to quantization by simulation. Simulation quantization implies further that any obstructions to quantization arise entirely within the quantum domain.
\end{abstract}

\maketitle

\section{Introduction}

\noi
Landauer's \cite{1} dictum ``information is physical'', implying that computers should be understood and studied as physical systems, marked computer science as a physical science.  The development of quantum computing and quantum information theory further consolidated this physical point of view and with this consolidation came the obverse inclination to view physics itself, or at least portions thereof, as a computer science.  Deutsch \cite{2} spells out this latter tendency in his influential paper on quantum computing and the physical interpretation of the Church-Turing thesis:
\begin{quote}
To view the Church-Turing hypothesis as a physical principle does not merely make computer science into a branch of physics.  It also makes part of experimental physics into a branch of computer science \dots The existence of a universal quantum computer \dots implies that there exists a program for each physical process.
\end{quote}
\noi
Smith \cite{3}, concurring, argues that the Church-Turing thesis can be
\begin{quote}
\dots interpreted as a profound claim about the physical laws of our universe, i.e.: {\em any physical system that purports to be a ``computer'' is not capable of any computational task that a Turing machine is incapable of}.
\end{quote}
And, in his  popular account of quantum computing, Aaronson \cite{4} metaphorically extends again the reach of computer science into physics,
\begin{quote}
Basically, {\it quantum mechanics is the operating system that other physical systems run on as application software} \dots There's even a word for taking a physical theory and porting it to this OS: ``to quantize.''
\end{quote}
Aaronson's metaphor, however, conceals a complication, the problem of (first) quantization, how to port classical systems to the quantum operating system. 
\par
First quantization is a mystery, or so Nelson opined.  The assertion that ``classical physics is false''\cite{2}, does not dispel the mystery, at least not at the level of elementary quantum mechanics.  Here, the logical relationship between classical and quantum systems is not well understood.  In fact, given that quantum mechanics postulates the quantum Hamiltonian (the fundamental quantum observable) shall be defined by quantizing a corresponding classical Hamiltonian \cite{5}, one could almost characterize the relationship as circular.  However, a more practical problem is that there are obstructions to quantization; that is, there is no universal quantization procedure.  Certain classical systems cannot be consistently quantized and thus cannot be ported to Aaronson's quantum OS.
\par
In this paper, I show that if we take Aaronson's metaphor literally, or, more specifically, if we assume that physical versions of the Church-Turing thesis are true, then we are faced with several dilemmas principal of which is that we can define a bijection between all simulable, i.e., finitely realizable, classical and quantum systems.  In other words, as far as computer science is concerned, classical and quantum physics are indistinguishable.  This first dilemma introduces a second: either there are intuitively simulable classical systems that are not Turing computable or, contrary to known mathematical results, there are no obstructions to quantization for simulable classical systems. What this latter alternative actually means depends, as we shall see, on what quantization means. But at the very least, it will entail abandoning the idea that quantization is about turning classical systems into quantum systems; that is, any obstructions to quantization are purely quantum obstructions.  
\par
Here, and in the following, I assume familiarity with the basic conceptual background of classical and quantum computing including the notions of Turing machine (TM), universal Turing machine (UTM), quantum computer, universal quantum computer, and the Church-Turing (CT) thesis (some references are \cite{6} and \cite{7}).  In the next section, I review some of this background primarily to set the stage for later constructions and to motivate the hybrid definition of simulation I give there.  In section 3, I discuss the quantization problem, obstructions to quantization associated with the Groenewold and van Hove no-go theorem, and define a new, simulation-based approach to quantization.  In section 4, I outline results on the definition, integrability, and quantization of nonlinear oscillators, and show, among other things, that if there exist obstructions to simulation-based quantization, then there exist intuitively simulable physical systems, the aforementioned nonlinear oscillators, that are not Turing computable. I conclude in section 5, by pointing out further implications of the physical versions of the CT thesis including a possible connection to recent tests of Bell's inequality and superdeterminism theories.  

\section{Physical Versions of the Church-Turing Thesis}

\subsection{Physical CT Thesis and Principle}

Originally, the CT thesis was a mathematical conjecture characterizing the computational capacities of human computers.  It can be stated as follows \cite{2}:
\par\medskip\noindent
{\bf CT Thesis}\quad Every function which would be naturally regarded as computable can be computed by the universal Turing machine.
\par\medskip
With the advent of computing machines as opposed to pencil-and-paper human computers, a broader, physical, machine-centric version of the CT thesis has taken hold \cite{8}:
\par\medskip\noindent
{\bf Physical CT Thesis}\quad A TM  can do (compute) anything a computer can do.
\par\medskip\noindent
This is a broader version of the CT thesis not only because it asserts Turing machines can compute what {\it machines}, e.g., laptops, can compute, but because it asserts that TMs can ``model {\it all} computation'' \cite{8}.  (Note, there is a ``strong CT thesis'' \cite{7} involving computational complexity.  It states that a TM can perform any computation with at most polynomial slowdown.  In this paper I am concerned with issues relating to simple computability.)
\par
Deutsch points out that although the original CT thesis is a ``quasi-mathematical conjecture \dots underlying this \dots hypothesis there is an implicit physical assertion''\cite{2}. Deutsch elevates this implicit physical assertion to an explicit physical principle:
\par\medskip\noindent
{\bf CT Principle}\quad Every finitely realizable physical system can be perfectly simulated by a universal model computing machine operating by finite means.
\par\medskip
Congruent with Landauer's dictum, Deutsch's CT principle proceeds from the intuition that a computer is a kind of physical system; ``a computing machine is any physical system whose dynamical evolution takes it from a set of `input' states to a set of `output' states''\cite{2}. At the same time, the principle appears to extend the reach of the physical CT thesis by positing that a physical system of the appropriate computational type can model, i.e., simulate, not only all computation, but the motions or dynamics of all finitely realizable systems (following Nielsen \cite{9}, I interpret this last phrase to mean systems constructible in a laboratory, computers themselves, for instance). 
\par
Much of the above was anticipated and colloquially summarized by Feynman \cite{10} in his groundbreaking paper on computer simulation and quantum computing:
\begin{quote}
What kind of computer are we going to use to simulate physics?  Computer theory has developed to the point where it realizes that it doesn't make any difference; when you get a {\em universal computer}, it doesn't matter how it's manufactured, how it's actually made.
\end{quote}
It is important to note, however, that in neither Feynman's quote nor in my discussion of the CT thesis and principle above is the meaning of simulation pinned down.  In the next section, I propose a remedy for this oversight.

\subsection{Simulation}

Both versions of the CT thesis, and the CT principle, are conjectures concerning simulation by computing device (possibly human); however, they invoke different notions of simulation. The original CT thesis asserts that all computable functions are computable by TM and all TMs are computable, meaning simulable, by a UTM, the latter by encoding the simulated TM and its input as a program in the UTM.  Deutsch's principle on the other hand refers to ``perfect simulation'' of physical systems by universal (quantum) computer meaning the perfect matching of the simulated system's input and output distributions by the simulating computer.  
\par
The manner in which a UTM simulates a TM is {\em intensive} in the sense that the internal (computational) behavior of the UTM is indistinguishable from the simulated TM.  It is this intensive, mimicry, view of simulation that Feynman has in mind in the previously quoted paper \cite{10} on simulating physics.
\begin{quote}
So what kind of simulation do I mean?  There is, of course, a kind of approximate simulation in which you design numerical algorithms for differential equations, and then use the computer to compute these algorithms and get an approximate view of what physics ought to do.  That is an interesting subject, but is not what I want to talk about.  I want to talk about the possibility that there is to be an {\em exact} simulation, that the computer will do {\em exactly} the same as nature.
\end{quote}
\par
Feynman's exact simulation is synonymous with intensive simulation.  In contrast, Deutsch's ``perfect'' simulation is {\em extensive}, inputs and outputs of the simulated system and simulating computer must agree in some way, but their internal behavior (``motions" \cite{2}) need not.  However, as Deutsch  \cite{11} later emphasizes, given any computer C, a universal computing machine, specifically, a network of universal quantum gates, ``could emulate more than just the relationship between the output of C and its input.  It could produce the output {\em by the same method} -- using the same quantum algorithm -- as C''. 
\par
There are further uncertainties in the literature concerning the meaning of simulation.  For instance, Piccinini \cite{12} argues that Deutsch's CT principle is ``ambiguous between two notions of simulation.''  On the one hand, it refers to simulation of one computing device by another, on the other, it entails computational modeling, ``the approximate description'' of a physical system's dynamics (Feynman's ``numerical algorithms for differential equations''). According to Piccinini, with respect to the first notion of simulation the CT principle is a form of ``ontic pancomputationalism'', the assertion that ``everything in the universe is a form of digital computing system''\cite{12}.  If instead the CT principle is understood with respect to approximate simulation, then it is ``the claim that any physical process can be computationally approximated to the degree of accuracy desired in any given case''\cite{12}.  
\par
For the purposes of this paper, I adopt the following hybrid definition of simulation:
\par\medskip\noindent
{\bf Definition (Simulation)}\quad A computing device simulates a physical system if it can intensively approximate the dynamical evolution and observables of that system to the degree of accuracy desired in any given case.
\par\medskip\noindent
Here, the phrase physical system includes computing devices; and intensive approximation includes the limiting case in which the simulating computing device behaves exactly, in Feynman's sense (or Turing's, in the classical theoretical case), like the simulated system.  My definition is perhaps closest to Sandberg and Bostrom's \cite{13} definition of emulation: a computer emulates another computer or system if it ``mimics the internal causal dynamics [of the emulated computer or system] (at some suitable level of description)''.   
\par
There are several reasons for defining simulation in this way: first, it reflects extant approaches to simulation in the literature; second, it leads to counterintuitive results; and third, because it leads to counterintuitive results it may provoke researchers to clarify what simulation by computer means. 

\section{Quantization}

\noi
``Classical physics is false''\cite{2}; quantum physics is true; so noted.  Still, in practice, virtually all (non-relativistic, non-spin) quantum systems are (first) quantizations of classical systems. As Giulini \cite{14} writes, 
\begin{quote}
So far a working hypothesis has been to {\em define} quantum theories as the results of some 'quantization procedures' after their application to classical theories. One says that the classical theory (of 'something') 'gets quantized' and that the result is the quantum theory (of that 'something').
\end{quote}
\par
In this section, I outline the quantization problem, that is, the problem of associating a quantum system to a given classical one. This problem is not solvable in general; there is no consistent scheme for quantizing all classical systems.  After reviewing certain obstructions to quantization, the no-go result of Groenewold and van Hove, I  propose a new, simulation approach to quantization of (simulable) classical systems based on the fact that quantum computers obey the physical CT thesis \cite{7}.  I also describe a second, presumptively stronger form of simulation quantization based on Deutsche's CT principle.  However, we will see that my hybrid definition of simulation allows me to identify these weak and strong forms of simulation quantization.

\subsection{The Quantization Problem}

In mathematical terms, the first, i.e., non-field-theoretic, quantization problem is to find a rule that simultaneously assigns to any symplectic phase space $P=(P,\omega)$, representing the states of a classical physical system, a complex Hilbert space ${\cal H}_P$, representing the states of a quantum system, and to any classical observable $f\in C^\infty(P)$, a smooth function on $P$, an operator in the vector space ${\cal A}({\cal H}_P)$ of self-adjoint operators on ${\cal H}_P$.  If one restricts attention, as I will here, to classical systems with Euclidean phase space $P=({\bb R}^{2n},\omega)$, a solution to the quantization problem is a representation 
\begin{align*} 
\rho:C^\infty(P)&\to {\cal A}({\cal H}_P)\\
f&\mapsto\rho(f)
\end{align*}
satisfying the following conditions:
\begin{enumerate}
\setlength{\itemindent}{\parindent}
\item $\rho(f+g)=\rho(f)+\rho(g)$,
\item $\rho(\lambda f)=\lambda\rho(f)$ for $\lambda\in\bb R$,
\item $\rho(\{f,g\})=(i/\hslash)\left[\rho(f),\rho(g)\right]$
\item $\rho(1)=I$, where 1 is the identity and $I$ the identity operator,
\item $\rho(q_i)$ and $\rho(p_i)$ act irreducibly on ${\cal H}_P$.
\end{enumerate}
A representation satisfying these five conditions is called a {\it full quantization} of $P$ with $(q_i,p_i)=(q,p)$ canonical coordinates on $T^*P\cong{\bb R}^{2n}$, $\{\cdot,\cdot\}$ the Poisson bracket on the Lie algebra $C^\infty(P)$, and $\left[\cdot,\cdot\right]$ the commutator on the Lie algebra ${\cal A}({\cal H}_P)$.  By the Stone-von Neumann theorem, condition 5 implies $\rho$ is the Schr\"odinger representation where ${\cal H}_P=L^2({\bb R}^n)$, $\rho(q)=q$, and $\rho(p)=(1/i)\partial/\partial{q}$ \cite{15}.  (I am glossing over a number of technical details.  For a more precise account consult \cite{14} or \cite{15}.)
\par
For $P={\bb R}^{2n}$, Groenewold \cite{16} and van Hove \cite{17} proved the following no-go theorem: 
\par\medskip\noindent
{\bf Theorem (Groenewold-van Hove)}\quad There exists no full quantization of the Lie subalgebra of polynomial observables $\cal P$ in $C^\infty(P)$.
\par\medskip\noindent
Hence there is no full quantization of $C^\infty(P)$.  At the same time, van Hove showed that restricting $\rho$ to the Heisenberg subalgebra ${\cal P}^1=\hbox{span} \{1,q,p\}$ of $\cal P$ produces the standard Schr\"odinger quantization of $P$.  (Note, the Heisenberg algebra is the minimal subalgebra in $\cal P$ that coordinatizes phase space $P$.)  Furthermore, $\rho$ can be extended to a full quantization of the Lie subalgebra ${\cal P}^2=\hbox{span}\{1,q,p,q^2,p^2,qp\}$.  Gotay \cite{18}  refines these results; for polynomial functions on ${\bb R}^{2n}$ he shows there are precisely two maximal Lie subalgebras, ${\cal P}^2$ and ${\cal P}^{\infty,1}$,  that contain the Heisenberg algebra ${\cal P}^1$ and are quantizable. (The Lie subalgebra ${\cal P}^{\infty,1}$ is the subalgebra of polynomials linear in $p$ with coefficients arbitrary polynomials in $q$.)  In short, there is no full quantization for polynomial Lie subalgebras on ${\bb R}^{2n}$ containing terms with degree greater than 2 in both $p$ and $q$.
\par
Although the Groenewold-van Hove no-go theorem appears special in that it applies to the flat phase spaces ${\bb R}^{2n}$, obstructions to quantization are generic in the following category-theoretic sense. Let $\mathfrak{C}$ denote Weinstein's \cite{19} classical category of symplectic manifolds and symplectomorphisms, and $\mathfrak{Q}$ the quantum category of complex Hilbert spaces and unitary transformations; then the quantization problem can be restated as the more general problem of finding a  functor $\Delta:\mathfrak{C}\to\mathfrak{Q}$ consistent with the Schr\"odinger quantization.  Gotay \cite{20}, utilizing Groenewold and van Hove's result, proves no such functor exists.

\subsection{Weak and Strong Simulation Quantization}

The physical CT thesis and the CT principle suggest an alternative approach to the quantization problem based on simulation of classical systems.  Rather than construct a quantum representation of a classical system, one solves the quantization problem by constructing a (universal) quantum computer that simulates the classical system.  This quantum simulation is then defined to be the quantization of the classical system.  As discussed below, the resulting simulation-based quantization has weak and strong forms corresponding to the physical CT thesis and CT principle, respectively, although we will also see that the hybrid definition of simulation allows us to identify these forms.
\par
Simulation-based quantization embodies the algorithmic conceptualization of quantization described by Berezin \cite{21}.
\begin{quote}
It is generally accepted that \dots quantization is an algorithm by means of which a quantum system corresponds to a classical dynamic one.  
\end{quote}
Berezin continues:
\begin{quote}
Furthermore, it is required that in the limit $h\to0$ where $h$ is the Planck's constant, a quantum dynamic system change[s] to a corresponding classical one.  This requirement is called the correspondence principle.  It is quite obvious that there exist quite a lot [of] quantizations obeying the correspondence principle; the quantum description of a physical phenomena is more detailed than the classical one, and so there are certain phenomena the difference between which is displayed in their quantum description, whereas their classical description does not show this difference.
\end{quote}
Below we will see how a truly algorithmic approach to quantization, based on physical versions of the CT thesis, leads to a very different relationship between classical and quantum systems compared to the conventional one Berezin describes.
\par
It may help to introduce some notation. (The reader is forewarned that rigor does not thereby ensue.  The following notation helps me clarify certain ideas, but the arguments remain  informal.)  Let $C_{\rm U}$ denote a universal classical computer corresponding to the physical instantiation of a universal Turing machine and let $ Q_{\rm U}$ be a universal quantum computer corresponding to the physical instantiation of a universal quantum computer (for example, $Q_{\rm U}$ could be a quantum computer constructed from a network of universal quantum gates).  Then weak simulation quantization is defined as a map
\begin{align*}
sq_w:{\cal C}_c&\longrightarrow{\cal Q}_c\\
x&\longmapsto Q_{\rm U}(x)
\end{align*}
where ${\cal C}_c$ is the set of all classical computers, ${\cal Q}_c$ the set of all quantum computers, and  $Q_{\rm U}(x)$ is the quantum simulation of the classical computer $x\in{\cal C}_c$.  (Again, here and in what follows, simulation means hybrid simulation as defined in section 2.2.)  Because quantum computers obey the physical CT thesis  \cite{7}, under the assumption that the physical CT thesis is true the quantum simulation $Q_U(x)$ defining $sq_w$ is guaranteed to exist for any $x\in{\cal C}_c$, and hence all classical computers are weak simulation quantizable.  
\par
In addition, given any quantum computer $y\in{\cal Q}_c$, the physical CT thesis guarantees the existence of a weak simulation dequantization or correspon-dence-principle map
\begin{align*}
sd_w:{\cal Q}_c&\longrightarrow{\cal C}_c\\
y&\longmapsto C_{\rm U}(y)
\end{align*}
constructed in the obvious way.  In the exact, hybrid simulation Feynman-Turing-limit, the simulations $sq_w$ and $sd_w$ are inverses:
\[
sd_w\circ sq_w=1_{{\cal C}_c}\quad\hbox{and}\quad sq_w\circ sd_w=1_{{\cal Q}_c}
\]
where $\circ$ is the composition of quantization and dequantization maps induced by the simulation compositions $C_U(Q_U(x))$ and $Q_U(C_U(y))$, respectively, and the identity maps indicate self-simulation, e.g., $1_{{\cal C}_c}(x)=x$: the self-simulation of classical computer $x$ is simply $x$ itself. In fact, for any fixed level of accuracy, the approximate simulations $sq_w, sd_w$ are also inverses resulting in an (infinite) family of quantization, dequantization correspondences.   
\par
The CT principle defines strong forms of simulation quantization and dequantization.  Let $P$ be a classical system ($P$, or its states, might  be mathematically realizable, for example, as a symplectic manifold, but this is not required). Strong simulation quantization is defined as a map
\begin{align*}
sq_s:{\cal C}&\longrightarrow{\cal Q}\\
P&\longmapsto Q_{\rm U}(P)
\end{align*}
where ${\cal C}$ is the set of all finitely realizable classical systems, ${\cal Q}$ the set of all finitely realizable quantum systems, and $Q_{\rm U}(P)$ is the quantum simulation of the classical system $P\in{\cal C}$.  The existence of $sq_s$ is guaranteed by the CT principle, which, as discussed earlier, extends the physical CT thesis to include simulation of all finitely realizable physical systems.
\par
As with weak simulation quantization, strong simulation quantization has an associated dequantization map
\begin{align*}
sd_s:{\cal Q}&\longrightarrow{\cal C}\\
H&\longmapsto C_{\rm U}(H)
\end{align*}
where $H\in{\cal Q}$ is a finitely realizable quantum system.  Analogous to the weak case, in the exact limit $sq_s$ and $sd_s$ are inverse mappings:
\[
sd_s\circ sq_s=1_{{\cal C}}\quad\hbox{and}\quad sq_s\circ sd_s=1_{{\cal Q}}
\]
with $\circ$ induced by simulation compositions and where the identity maps are defined on all finitely realizable classical and quantum systems.  As in the weak case, there exists a family of strong quantization, dequantization correspondences parametrized by degrees of accuracy.
\par
Clearly, ${\cal C}_c\subseteq\cal C$ and ${\cal Q}_c\subseteq\cal Q$, hence strong entails weak simulation quantization.  However, in the exact, Feynman-Turing limit of intensive hybrid simulation we can identify a finitely realizable classical physical system with a subprocess (program) of its universal simulator (classical or quantum), hence in the exact limit $sq_w$ entails $sq_s$.  In this case, it makes sense to simply refer to simulation quantization and dequantization maps
\[
sq:{\cal C}\longrightarrow{\cal Q}\quad\hbox{and}\quad sd:{\cal Q}\longrightarrow{\cal C}
\] 
without the ``weak'' or ``strong'' qualifiers. I adopt this abbreviated terminology below.
\par
To recap, simulation quantization solves the quantization problem by constructing a program running on a quantum computer that reproduces the entire behavior, dynamics and observables, of the simulated classical system with any desired level of accuracy.   This is a truly algorithmic, indeed physical, approach to quantization unlike standard methods, which produce theoretical representations of the classical system and are provably not algorithmic.  
\par
Simulation quantization differs from standard quantization in another important respect; it has an inverse, simulation dequantization.  Although this property is not subject to mathematical proof, it is, as indicated above, an informal consequence of the definition of simulation.  By construction, there is (at any fixed degree of accuracy on in the exact simulation limit) one correct (real) quantum simulation of any given finitely realizable classical system since the given (real) system is unique; hence there is one correct simulation quantization of this system.  Similarly, there is one correct classical simulation of any finitely realizable quantum system because this system is likewise unique, hence there is one correct simulation dequantization of this system.  The resulting bijection
\[
\mathcal{C}\cong\mathcal{Q}
\] 
between finitely realizable classical and quantum systems can be interpreted to mean that computer science, through its acceptance of physical versions of the CT thesis and exact, intensive simulation, cannot computationally distinguish between classical and quantum physics.  (This uniqueness argument seems reasonable metaphysics, but see for example the conflicting metaphysics associated with homotopy type theory~\cite{22}, Leibniz equivalence~\cite{23}, and mathematical structuralism~\cite{24}.)
\par
A further implication is that in this restricted computer-science universe, there are no obstructions to quantization: every finitely realizable classical system is simulation quantizable.  This conclusion is unproblematic if all finitely realizable classical systems have standard quantizations, in which case simulation quantization does not contradict standard obstruction results.  However, as described in the next section, there appear to be finitely realizable systems, integrable nonlinear Hamiltonian oscillators, that are simulation quantizable, but are not quantizable in the usual sense.  (Note, from now on the phrase ``obstructions to quantization''  will mean the system in question is subject to the Groenewold and van Hove no-go theorem or some generalization thereof.)
\par
It might be useful at this juncture to reexamine Berezin's ``generally accepted'' notions about quantization in light of the definition and properties of simulation quantization.  As already mentioned, simulation quantization is a quantization algorithm, physically instantiated.  First quantization is a mystery because it is not an algorithm; this, in essence, is the meaning of the Groenewold-van Hove no-go theorem.  The powerful principle that allows us to construct a quantization algorithm, the CT principle (the physical CT thesis will do), also allows us to construct the inverse correspondence-principle or dequantization algorithm.  Thus, according to the CT principle, and contradicting Berezin, the quantum description of a (finitely realizable) physical phenomenon is {\em not} more detailed than the classical one.  In fact, even the correspondence limit $h\to0$ no longer appears, though, conjecturally, it might reappear in the intensive, hybrid simulation limit.  As discussed in more detail below, simulation quantization relocates standard quantization schemes and the correspondence principle entirely within either the quantum (or classical) domain. 
\par
Others do not drawn the same conclusions regarding the physical implications of physical interpretations of the CT thesis.  Deutsch provides one explanation for this, based on what might be called the ``too strong'' or continuum exception.  In effect, Deutsch situates classical, that is, continuous physics, outside of computer science.  The CT principle, he writes [2],
\begin{quote}
\dots is so strong that it is {\em not} satisfied by Turing's machine in classical physics.  Owing to the continuity of classical dynamics, the possible states of a classical system necessarily form a continuum.  Yet there are only countably many ways of preparing a finite input for $\cal T$ [UTM].  Consequently $\cal T$ cannot perfectly simulate any classical dynamical system.
\end{quote} 
Like the Chesire cat all that remains of the classical computer is its countable grin.  But then, what sort of device is the computer on which I am writing this sentence?  If it is not a classical dynamical system, then what is it?  My computer, modulo its quantum aspects, its laser hard drive, etc., is a finitely realizable physical system and is self-simulating as I type.  While the continuum problem is a barrier to understanding the boundary between classical and quantum systems, the continuum exception disembodies classical computing altogether.  Turing's {\em machine} becomes a ghost; information is no longer physical.  Perhaps more significantly, as ``no experiment allows us to have access to real phenomena on a very small scale''\cite{25}, the continuum exception has no real empirical content. 
\par
Of course, one could go one better and simply deny that classical computers as classical dynamical systems exist. For, as noted earlier, classical physics is false.  An honest if blunt philosophy, this implies, like Deutsch's continuum exception, that Turing machines model nothing (at least, nothing classical; according to the Church-Turing principle, they model, that is, simulate all finitely realizable quantum systems). It also renders moot a variety of heretofore interesting questions in computer science and computational complexity theory, not least of which is the question concerning the relative efficiency of quantum versus classical computers.  More importantly, denying that classical dynamical systems exist denies the mysterious relationship between quantum and classical physics and therefore between quantum and classical computers, and thus is unlikely to be the last word on the subject.
\par
The CT principle is, as Deutsch says, a strong principle.  Quantum and dequantum (classical) algorithms are actually beside the point; there is simply a universal machine.  The issue taken up next is how this universal machine, or to avoid being accused of ontic pancomputationalism, this computer-science universe of finitely realizable systems, relates to the universe of physics as understood by physicists. 

\section{Nonlinear Oscillators}

\noi
In this section, I describe an integrable, in fact solvable, nonlinear oscillator ``manufactured'' by Bruschi and Calogero \cite{26}.  Because it is integrable and solvable, this oscillator is presumably simulable, or finitely realizable, and hence simulation quantizable.  However, this classical system fails to be quantizable in the usual sense since its Hamiltonian is of normal type and quartic in the canonical coordinates $q$.  
\par
The following background discussion of many-body Hamiltonian mechanics and nonlinear oscillators is from Calogero \cite{27} and Bruschi and Calogero \cite{26}.  Further details can be found in these sources and in \cite{28} and \cite{29}.

\subsection{Integrable $N$-body Hamiltonian Systems}

An $N$-body Hamiltonian system in canonical coordinates $q_i$ and momenta $p_i$, $i=1,\ldots,N$, is characterized by a Hamiltonian function $H(q_i,p_i)$ satisfying the following equations of motion:
\begin{align*} 
\dot q_i&=\partial H(q,p)/\partial p_i,\\
\dot p_i&=-\partial H(q,p)/\partial q_i.
\end{align*}
Here an overdot indicates differentiation with respect to time $t$ and $q=(q_1,\ldots,q_N)$, $p=(p_1,\ldots,p_N)$ are $N$-vectors.
\par
The Poisson bracket $\{\cdot,\cdot\}$ on system observables $f(p,q)$ and $g(p,q)$ is defined by
\[
\{f,q\}=\sum_{i=1}^N\left({{\partial f}\over{\partial q_i}}{{\partial g}\over{\partial p_i}}-
{{\partial f}\over{\partial p_i}}{{\partial g}\over{\partial q_i}}\right),
\]
with the time evolution of any observable quantity $f(p,q)$ given by
\[
\dot f=\{f,H\}.
\]
From this last equation it follows that an observable $f(p,q)$ that commutes with $H$, $\{f(p,q),H(p,q)\}=0$, is a constant of the motion,
\[
\dot f(p,q)=0.
\]
We say an $N$-body Hamiltonian system is (Liouville) integrable if it has $N$ observables $f_i(p,q)$, $i=1,\ldots,N$, that are constants of the motion and are in involution, meaning,
\[
\{f_i(p,q),f_j(p,q)\}=0,
\]
for $i,j=1,\ldots N$.
\par
All one dimensional Hamiltonian systems are integrable since the Hamiltonian itself satisfies $\{H(p,q),H(p,q)\}=0$.  However, integrable systems are special since their time evolution cannot be chaotic, and ``chaotic behavior is in some sense {\it generic} for Hamiltonian systems with confined motions \dots'' \cite{27}.  Even if the motion of an integrable system is regular, it may still be very difficult to find solutions to its equations of motion.  Integrable systems for which solutions can be found, or can be determined using purely algebraic operations, are called solvable.
\par
One further piece of terminology, a Hamiltonian system is of normal type if its Hamiltonian has the form
\[
H(p,q)=T(p)+V(q),
\]
where
\[
T(p)={1\over2}\sum_{i=1}^Np^2/m_i
\]
is a kinetic energy term in the canonical momenta with point masses $m_i$ and $V(q)$ is a potential function of the canonical coordinates, independent of the momenta.

\subsection{Nonlinear Oscillators}

Calogero \cite{27} (based on work first reported in \cite{26}) shows how to manufacture integrable Hamiltonian systems from exactly treatable, matrix evolution equations.  This is a three-step process, which I outline in the case of the (quartic) nonlinear oscillator of special interest here.  
\par
First Calogero proves that the matrix equation 
\begin{equation}
\ddot U = {1\over2}\left(AU+UA\right)+bU^3,\label{1}
\end{equation}
with $U$ an $N\times N$ matrix of arbitrary rank, $A$ a constant matrix, and $b$ a scalar, is integrable, indeed solvable, as a special, periodic case of the non-abelian Toda lattice, a system shown solvable by Krichever \cite{30}.  Equation (1) can also be expressed in terms of Lax pairs and hence is integrable for this reason as well \cite{27}. 
\par
Second, Calogero parametrizes the matrix evolution equation (1) in terms of 3-vectors, producing Newtonian evolution equations.  This last, third step creates a number of different Newtonian equations of motion each interpretable as many-body, normal-type Hamiltonian systems in three-dimensional Euclidean space.  These systems are velocity-independent and invariant under rotations of 3-space and hence can be viewed as systems of nonlinear oscillators.  Indeed, one of the systems constructed in this manner is an anharmonic oscillator whose Hamiltonian potential term is quartic in the canonical coordinates (equation 3.3 in \cite{26}):
\begin{IEEEeqnarray*}{rCl}
H& =&\;{\frac12}\sum^N_{i,j=1}\left[\left(\vec{p}_{ij}\cdot\vec{p}_{ji}\right)-\pi_{ij}\pi_{ji}\right]
-{\frac12}\sum^N_{i,j,k=1}a_{ij}\left[\left(\vec{r}_{jk}\cdot\vec{r}_{ki}\right)-\rho_{jk}\rho_{ki}\right]\\
& &-{\frac b4}\sum^N_{i,j,k,l=1}\big\{
2\left[\rho_{ij}\rho_{kl}\left(\vec{r}_{jk}\cdot\vec{r}_{li}\right)
+\rho_{ij}\rho_{li}\left(\vec{r}_{jk}\cdot\vec{r}_{kl}\right)
+\rho_{ij}\rho_{jk}\left(\vec{r}_{kl}\cdot\vec{r}_{li}\right)
\right]	\\
&&-\rho_{ij}\rho_{jk}\rho_{kl}\rho_{ki}+2\left[\left(\vec{r}_{ij}\cdot\vec{r}_{kl}\right)
\cdot\left(\vec{r}_{jk}\cdot\vec{r}_{li}\right)
-\left(\vec{r}_{ij}\cdot\vec{r}_{kl}\right)
\cdot\left(\vec{r}_{jk}\cdot\vec{r}_{li}\right)\right.\\
&&-\left.\left(\vec{r}_{ij}\cdot\vec{r}_{kl}\right)
\cdot\left(\vec{r}_{jk}\cdot\vec{r}_{li}\right)\right]
-\left(\vec{r}_{kl}\wedge\vec{r}_{li}\right)\cdot\left(\rho_{ij}\vec{r}_{jk}+\rho_{jk}\vec{r}_{ij}\right)\big\}
\end{IEEEeqnarray*}
In this equation, the canonical coordinates are expressed as 3-vectors $\vec{r}_{nm}$ and pseudo-scalars $\rho_{nm}$; the canonical momenta as 3-vectors $\vec{p}_{nm}$ and pseudo-scalars $\pi_{nm}$ with $n,m=1,\ldots,N$.  The operators $\wedge$ and $\cdot$ are the standard wedge and scalar products on 3-vectors.  
\par
As Bruschi and Calogero \cite{26} observe, the nonlinear oscillator defined by $H$ has
\begin{quote}
\dots obvious and ubiquitous applicative interest, inasmuch as it generally provides the first nonlinear (`unharmonic') correction to the behavior of linear (`harmonic') oscillators, the physical relevance of which is of course universal.
\end{quote} 
Furthermore, being both integrable and solvable, it seems reasonable to conclude that the dynamical system defined by $H$ is simulable, and is therefore simulation quantizable.  However, $H$ is quartic in the 3-vectors $\vec{r}_{nm}$, hence in the canonical coordinates, hence by Groenewold and van Hove's obstruction result (see section 3.1) the system defined by $H$ is not quantizable.  The Hamiltonian $H$ defines a relatively simple classical system that is simulation quantizable, but is not quantizable in the usual sense.
\par
If this result is surprising, the alternative is no less so.  If there is an obstruction to the simulation quantization of Bruschi and Calogero's anharmonic oscillator, then we have discovered a regular classical system of ``applicative interest,'' the first nonlinear correction to the ubiquitous harmonic oscillator, that is not quantum simulable and hence is not classically simulable.  In short, $H$ defines a simulable dynamical system that is not Turing computable.

\subsection{Quantizing Simple Nonlinear Oscillators}

It is beyond the scope of this paper (to attempt) to quantize, in the standard sense, Calogero and Bruschi's quartic oscillator.   Still, some of the peculiarities one might expect from such an undertaking can be seen in Calogero and Graffi's \cite{31} and Calogero's \cite{32} quantization of three simpler, one degree of freedom nonlinear Hamiltonian oscillators.  This triumvirate of related systems are defined by the following classical Hamiltonians:
\begin{IEEEeqnarray*}{rCl}
H(p,q)&=&{\frac12}\bigg[{\frac{p^2q^3}c+c\bigg(q+{\frac1q}\bigg)}\bigg],\IEEEyesnumber\\
\IEEEnonumber\\
H(p,q)&=&{\frac12}\bigg[{\frac{p^2\sin^2(q)\sin(2q)}{2c}}+{\frac{2c}{\sin(2q)}}\bigg],\IEEEyesnumber\\
\IEEEnonumber\\
H(p,q)&=&{\frac12}\bigg[{\frac{p^2\sin^2(q)\sin(2q)}{2c}}+2c\cot(2q)\bigg].\IEEEyesnumber
\end{IEEEeqnarray*}
The constant $c$ in the Hamiltonian of equation (2), is a positive real number, while in equations (3) and (4), $c$ is an arbitrary real number.  Because the dynamical systems modeled by these Hamiltonians are isochronous, that is, all of their classical motions are nonsingular and periodic with period $2\pi$, it is ``natural'' \cite{31} to view them as integrable, nonlinear, unharmonic oscillators.
\par
It turns out that canonical (Weyl) quantization of these systems produces ``remarkable ambiguities'' \cite{31}.  The most relevant of these for highlighting the differences between simulation and standard quantization is the fact that for each the canonical quantization depends on the value of $c$, but the classical dynamics does not.  
\par
For example, the constant $c$ does not appear in the Newtonian equation
\[
\ddot{q}={\frac{3\dot{q}^2}{2q}}+{\frac12}q\left(1-q^2\right)\tag{5}
\]
associated to Hamiltonian (2) above.  This means that the canonical coordinate $q(t)$ does not depend on $c$ and that $c$ is a simple scale factor in the canonical momentum $p(t)$.  However, as Calogero and Graffi  demonstrate, the ground state $E_0$ of the quantized Hamiltonian does depend on $c$.  This peculiar result illustrates the well-known fact that, in general, canonical transformations do not commute with quantization \cite{33}; that is, ``\dots it may make a difference whether one quantizes before or after having performed a canonical transformation'' \cite{31}.  Calogero and Graffi go on to prove that one can eliminate all such ``\dots difficulties and paradoxes: at a cost, however, of imposing restrictions on the constant $c$ which have no classical component'' \cite{31}.  Lest the reader miss the point, they add:
\begin{quote}
The fact that these different alternatives depend on the value of the constant $c$, whose presence in the Hamiltonian (1) [(2) above] does not affect at all the behavior of the corresponding classical system \dots provides an example---striking by its remarkable simplicity---of the peculiarity of quantization.
\end{quote}
\par
Perhaps equally striking is the fact that simulation quantization can resolve these difficulties and paradoxes as well.  Assuming one-degree-of-freedom, integrable, isochronous oscillators are simulable, then they must be simulation quantizable via a unique quantum simulation, that is, one having a single ground state independent of the value of the constant $c$ in the classical Hamiltonian.  Berezin's observation (quoted in section 3.2) that there are phenomena distinguishable by their quantum, but not their classical description, anticipates these anharmonic systems exactly, and is completely wrong in the context of simulation quantization.  
\par
As in the case of the quartic oscillator, there is a converse and equally counterintuitive scenario.  If there is an obstruction to simulation quantization, then there exist simple one-dimensional classical systems that are intuitively simulable, but are not Turing computable.

\subsection{The Quantization Problem Revisited}

I argue above that there is a tradeoff between simulation and quantization for certain classical systems.  This tradeoff is unproblematic if we believe that the classical systems in question need not be simulable, or, alternatively, if simulation quantization does not solve the full quantization problem outlined in section 3.1.  Because the anharmonic oscillators described earlier are integrable they are presumably simulable.  Although I offer no proof of this assertion, it seems both intuitively reasonable and consistent with the literature (for example, on the non-computability of non-integrable systems, see \cite{34} or \cite{35}).
\par
Does simulation quantization solve the quantization problem? If quantization is broadly defined as ``an algorithm by means of which a quantum system corresponds to a classical dynamical one''\cite{21}, then on the informal strength of physical versions of the CT thesis it does.  However, because a simulating quantum system behaves exactly like the simulated classical system, the relationship between simulation quantization and full quantization as presented in section 3.1 may be less clear. 
\par
The fundamental conceptual difficulty seems to arise with the customary distinction between macro- and microscopic physical systems and their mathematical descriptions. As Guillemin and Sternberg \cite{36} write,
\begin{quote}
A\dots physical system usually has both a ``classical'' description which provides one with information about its macroscopic behavior ($\hbar$ small) and a ``quantum'' description which provides information about its microscopic behavior ($\hbar=1$).
\end{quote}
Vogan \cite{37} puts it this way, 
\begin{quote}
Even though classical and quantum mechanics are different, they can sometimes be regarded as different descriptions of ``the same'' physical system. That is, for each classical mechanical system there should be a corresponding quantum mechanical system system. 
\end{quote}
\par
What is startling about simulation quantization is that it implies microscopic systems behave exactly like macroscopic systems and vice versa. Simulation quantization produces interchangeable quantum and classical realizations of the same underlying physical system. (Indeed, this is the import of the bijection $\mathcal{C}\cong\mathcal{Q}$ between finitely realizable classical and quantum systems described in section 3.2.) Since simulation quantization exactly reproduces the entire dynamical behavior of the simulated classical system, it is a full (physical) quantization of that system, that is, one that reproduces every classical observable in quantum form (a quantum theory such as Ghirardi, Rimini, and Weber's spontaneous collapse theory \cite{38} may help to make intuitive sense of this full quantization, but is not required). Simulation quantization accomplishes more than is expected of full quantization as it obliterates the distinction between quantum and classical physics altogether. According to simulation quantization, classical and quantum simulations are different descriptions of the same (without Vogan's scare quotes) physical system.
\par
Simulation quantization is, so to speak, a hardware-level quantization of the simulated classical system. Presumably, this hardware-level (physical) quantization induces a software-level map between the mathematical models of the classical system and the simulating quantum system. Again, with no pretense to mathematical rigor, let $C$ denote a finitely realizable classical system and $Q_\mathrm{U}(C)$ its simulation quantization, $sq:C\to Q_\mathrm{U}(C)$. Under the assumption that $C$ can be mathematically modeled by a symplectic phase space $(P_C,\omega_C)$, and $Q_\mathrm{U}(C)$ by a Hilbert space $\mathcal{H}_{Q_\mathrm{U}(C)}$, simulation quantization induces a mathematical (software-level) full-quantization map $\rho_\mathrm{sq}$ defined by the following commutative diagram
\[\begin{tikzpicture}[scale = 6, node distance = 3cm, auto]
  \node (Pc) {$C$};
  \node (Qu) [right of=Pc] {$Q_{\rm U}\left(C\right)$};
  \node (Hq) [above of=Qu] {$\mathcal{A}\left(\mathcal{H}_{Q_\mathrm{U}(C)}\right)$};
  \node (Pw) [above of=Pc] {$C^\infty\left(P_C\right)$};
  \draw[->] (Pc) to node {$sq$} (Qu);
  \draw[<->] (Pc) to node {$\varphi_C$} (Pw);
  \draw[<->] (Qu) to node [swap] {$\varphi_{\tiny Q}$} (Hq);
  \draw[->][dashed] (Pw) to node {$\rho_{sq}$} (Hq);
\end{tikzpicture}\]
Here, $C^\infty(P_C)$ stands for the algebra of classical observables on the phase space $P_C$; $\mathcal{A}\left(\mathcal{H}_{Q_\mathrm{U}(C)}\right)$, the corresponding algebra of self-adjoint operators on the Hilbert space $\mathcal{H}_{Q_\mathrm{U}(C)}$; and $\varphi_C$, $\varphi_Q$ are maps associating the usual mathematical models to the physical systems $C$ and $Q_\mathrm{U}\left(C\right)$, respectively. Note, the existence of these modeling maps $\varphi_C$ and $\varphi_Q$ has nothing to do with quantization, per se. The maps signify only the possibility of finding a mathematical description of the given classical and quantum systems. In particular, while it may be difficult to work out the quantum mechanical model implied by $\varphi_Q$, unlike the case with conventional mathematical approaches to full quantization, there is no known theoretical obstruction to doing so. 
\par
Whether or not one can determine the full quantization map $\rho_\mathrm{sq}$, simulation quantization based on the CT principle fundamentally alters the quantization problem by locating it entirely within the quantum domain. This means any mathematical obstructions to full quantization are now obstructions to producing the Hilbert space representation, the image of the modeling map $\varphi_Q$, associated with a known quantum system, the quantum simulation produced by simulation quantization. It is perhaps also worth noting that the mathematical, full-quantization problem can be pulled back to the classical domain: take a given quantum system, apply simulation dequantization, determine the mathematical description of the resulting classical system, and then return to the quantum domain via simulation quantization. Presumably even quantum systems with spin degrees of freedom can be quantized in this way.
\par
As mentioned earlier, physical versions of the CT thesis would likely change our understanding of the correspondence principle.  Because every finitely realizable classical system has a unique quantum simulation, an alternative correspondence principle arises in which quantum systems correspond to other quantum systems, the simulations of classical systems. Classical physics emerges then not as Planck's constant $h\to0$, but rather through dequantization of the intensive simulation limit of an entirely quantum correspondence.

\section{Conclusion}

\noi
How is the quantum simulation of a classical system related to its quantization?  Through informal arguments based on a hybrid definition of simulation and physical versions of the CT thesis, I have argued two main points. First, if physical versions of the CT thesis are true, then there exist simulation quantization and dequantization maps that define a one-to-one correspondence $\mathcal{C}\cong\mathcal{Q}$ between finitely realizable classical and quantum systems.  
\par
Second, either there are no obstructions to simulation quantization or there exist intuitively simulable classical systems that are not Turing computable. The correctness of traditional quantization methods are called into question if there are no obstructions to simulation quantization, while the alternative implies the existence of non-chaotic, incomputability.  This tradeoff underscores a broader issue: how is the universe studied by computer scientists, as characterized by the correspondence $\mathcal{C}\cong\mathcal{Q}$, related to the universe studied by physicists?  We have come, it appears, full circle, as this is precisely the sort of question the CT principle is supposed to address. 
\par
To close, I outline two additional results accruing to this general line of reasoning. The first may have practical consequences. The second relates to dequantization and concerns a longstanding issue in quantum theory.  
\par
As mentioned earlier, given a specified level of accuracy, hybrid simulation allows us to identify the sets ${\cal C}_c=\cal C$ and ${\cal Q}_c=\cal Q$, that is, respectively, the set of classical computers with the set of finitely realizable classical systems, and the set of quantum computers with the set of finitely realizable quantum systems.  One way of interpreting this result is that every computation, every simulation, {\em is} a physical system.  Computation thus reified is either physically meaningful or not.  In particular, a classical simulation is physically meaningless if the associated classical system is not quantizable.  The practical side of this arises as follows \cite{39}.  Suppose we are simulating a complicated, classical system and are able to make continuing improvements to the accuracy of our simulation.  Can we be sure that the simulation is converging to the system we are modeling, that is, can we be sure we are simulating what we think we are simulating?  If there are obstructions to simulation quantization and the physical system corresponding to our simulation is not quantizable, then the answer is no.  Indeed, the more accurate our classical simulation becomes the less we in some sense know about the system to which, we theorize, it converges.
\par
The proceeding focuses primarily on the quantization aspect of simulation and physical interpretations of the CT thesis. However, the consequences of the existence of the inverse, dequantization map are also curious.  As argued in section 3.2, the existence of the dequantization map follows from the CT principle; specifically, every finitely realizable quantum system can be simulated by a UTM.  By definition, UTMs are classical, deterministic, and local, the last because a UTM accesses information only on the portion of tape being scanned \cite{40}.  In other words,  a UTM can be viewed as a local hidden variable theory.  Recent tests of local realism have produced essentially loophole-free violations of Bell's inequality \cite{41}. In light of these results, the CT principle appears to connect UTM simulation with superdeterminism approaches \cite{42} to Bell's theorem. 

\vspace{0pt}
\bibliographystyle{amsplain}

\end{document}